# An Analytical Theory for the Early Stage of the Development of Hurricanes: Part II

By

Chanh Q. Kieu

Department of Meteorology, University of Maryland,

College Park, Maryland

July 2004

# Table of Contents

# Abstract

In this series of papers, an analytical theory for the early stage of hurricane development is proposed. In Part I, linear and non-linear theories have been formulated, based on an assumption of the existence of a positive feedback of a self-induced developing system. It was found in Part I that the linear theory, a kind of the 2D Rankine vortex, gives some unacceptable analytical solutions of hurricane development at the upper half of the atmosphere, although these solutions are fairly reasonable at the lower atmosphere. In the non-linear theory, the analytical solutions agree well in many respects with observations of hurricane development at the early stage. Particularly, the solutions of the non-linear theory offer some new insights, which cannot be obtained by the linear theory, such as the very fast increase of tangential wind near the core region, the deep U-shape of geopotential penetrating nearly up to the tropopause, the cyclonic wind at the upper atmosphere where divergence occurs. However, this non-linear has a disturbing property: a discontinuity of tangential wind and geopotential with radius as hurricanes develop. In this paper, this discontinuity will be resolved satisfactorily without any significant influence on the results obtained in the nonlinear theory in Part I. In addition, the effect of friction will be also investigated in this paper, which has not been done in Part I so far.

# Acknowledgment

I would like to thank Dr. Dalin Zhang for his many helpful corrections. I am also grateful to Mr. Wallace Hogsett and Emily Becker for their constructive comments. This research was supported by Vietnam Education Foundation (VEF). Had it not been for their support, I would not have been able to concentrate my mind on this work.

# 1. Introduction

A completely analytical theory for hurricane phenomenon, so far, has not been developed. Traditional theoretical approaches usually utilized some scaling numbers or incipient balanced relationships to study the hurricane development. For example, Sawyer (1947) and Yanai (1961) proposed an interesting theoretical idea for hurricane development, in which some initial equilibrium relationships between mass and wind fields were exploited (Palmen and Newton 1969, chapter 15). If this equilibrium is disturbed, tangential and radial wind will ensue. By following this route, Yanai has obtained a final equation which contains several stability coefficients. Because these coefficients have no explicit expression, no exact solution was obtained thus far. In another fashion, Willoughby (1979) has made use of some scaling values, e.g. tangential wind V=50m/s, radial wind U=10m/s, etc., and expanded all field variables as a series in terms of a nondimension number $\varepsilon$ ($\varepsilon \equiv U/V$). By performing some scaling and restricting the series to the first two approximations, a system of equations for mean fields and another system for perturbations emerged. This theoretical idea has employed implicitly a pre-existing hurricane and it was also impossible to find any analytical solutions with this theory because of the high order and mixed derivatives of field variables. Another example is the work of Montgomery and Farrell (1993) on tropical cyclone formation. In their work, a separation between geostrophic wind ($v_g$) and ageostrophic winds ($u_a$,w) was made to concentrate on the secondary circulation by ageostrophic winds. Some conclusions of the role of the upper level potential vorticity on the formation process of tropical cyclone were obtained but, once again, it is still impossible to come up with any exact solutions in their work. In recent decades, the most extensive approach to hurricane problem is modeling, due to the rapid improvement of computer speed (Zhang *et. al.* 2000; Challa *et. al.* 1998; Nguyen *et. al.* 2001; Zhu *et. al.* 2001; to name only a few). By using hurricane models, it is automatically impossible to find any information about the functional form of the solutions of the system of the primitive equations.

In Part I of this series of papers, two theories for the early stage of the development of hurricanes have been considered. One is a linear theory and the other is non-linear. In order to handle the problem analytically, it has been assumed that there exists a positive feedback process between latent heat and vertical motion in some region when (and where) a hurricane is expected to form. In the linear theory, all nonlinear terms are neglected and no friction included. The analytical solutions obtained from this theory have several fairly reasonable properties, *which agree with observations at the lower half of the atmosphere*, such as the correct variation of tangential wind with radius, the cyclonic sense of wind, the U-shape of geopotential (but shallower than observations (fig.2b and fig. 3 in Part I)), the increase of tangential and radial winds with time (which tell us about the growing of hurricanes as expected). These solutions are somewhat the same as for the 2D Rankine vortex in fluid mechanics. However, at the upper half of the atmosphere, this linear theory immediately gives us some unobserved behaviors: motion will be anti-cyclonic, a high pressure system (a hill-shape of geopotential) will develop.

To surmount these shortcomings of the linear theory, a nonlinear theory is considered (but it still has no friction included). This non-linear theory captures all the good points in the linear theory mentioned above (as it must). In addition, this theory offers some new properties. First, the tangential wind in the region where the positive feedback is effective will no longer increase as an exponential of time. Instead, it increases much stronger, as an exponential function of exponential of time ($\exp(e^{kt})$), which is much faster than the radial wind or the tangential wind outside this feedback region. This may help us explain why hurricanes have an eye formed in this region at later time (it does not mean that hurricanes have an eye at the early stage. It just means that hurricanes tend to form an eye). Second, this nonlinear theory gives us a correct variation of tangential wind with height as observed. Finally, this theory is also able to explain a potentially deeper U-shape of geopotential as well as the cyclonic motion even when there is a divergence in the upper half of the atmosphere, which was not obtained by the linear theory. The weakness of this non-linear theory is that tangential wind and geopotential are no longer a continuous function with radius as hurricanes develop. Another weakness, as in the linear theory, is the endlessly increase of solutions with time. This is because, so far, there is no mechanism to control the energy released in these theories. Both of these theories are expected to be correct up to first 6-12h of the growing of hurricanes. Beyond this period, they will give inaccurate descriptions.

The purpose of this paper will consist of:

1. Eliminate the discontinuity in the non-linear theory.

2. Include the effects of friction to the non-linear theory

This paper has five sections. Section 2 contains an improvement of the nonlinear theory in Part I so that tangential wind and geopotential will vary continuously with radius. Section 3 presents a nonlinear theory in which friction will be included. So far, this friction has been neglected in both theories in Part I. Section 4 will mention some unresolved questions. Discussions and conclusion will be given in the final section.

## 2. A continuous nonlinear theory for the early stage of hurricane development

Because of the remarkable results obtained from the non-linear theory in Part I by using an assumption of a positive feedback process within a region of radius **a**, this assumption will be made again. According to this assumption, every time hurricanes tend to be formed in some region, it is reasonable to assume that there is a proportional relationship between latent heat release (J) and vertical motion (w), given by: J=kw. More explanations about this proportionality are given in Part I and will not be mentioned here again. However, unlike the top-hat function of positive feedback process as in Part I, our purpose now is to find a new smooth function for the positive feedback process. Recall that, in Part I, vertical motion is assumed to be different from zero within a region of radius **a** and to be zero outside this region. Accordingly, the positive feedback process will vary with radius in the same way as the vertical motion. The natural generalization of a top-hat function is the Gauss-distribution function. This Gauss function appears frequently in many physical applications and it seems to be a good candidate for any "pulse" phenomenon. A hurricane can be regarded as a source point of some meteorological quantities, e.g. temperature or vertical motion, in some small scale region. These quantities then decrease very fast outward. The successes of the top-hat function of vertical motion in the non-linear theory in Part I suggest an attempt of using this Gauss function for the vertical motion to overcome the discontinuity.

For the sake of completeness, all assumptions in the nonlinear theory in Part I are rewritten here. More complete explanations can be found in sections 3.1 and 3.2 in Part I.

- Non-linear terms in the tangential and radial momentum equations are included, but non-linear terms *in vertical and thermodynamic equations* are neglected.

- No friction is considered

- Incompressible fluid $d\rho/dt=0$

- Boussisnesq approximation

- There exist a region in which vertical velocity (w) is related to heating source (J) by a positive feedback relationship (J=kw). Vertical velocity will have a prescribed Gauss profile with radius.

- F-plane

- Hurricanes are axis-symmetric

We first start with the system of the primitive equations in the cylindrical coordinate:

$$\frac{\partial u}{\partial t} + u\frac{\partial u}{\partial r} + w\frac{\partial u}{\partial z} - \frac{v^2}{r} = -\frac{\partial \phi}{\partial r} + fv \qquad (1.1)$$

$$\frac{\partial v}{\partial t} + u\frac{\partial v}{\partial r} + w\frac{\partial v}{\partial z} + \frac{uv}{r} = -fu \qquad (1.2)$$

$$\frac{\partial w}{\partial t} = b \qquad (1.3)$$

$$\frac{1}{r}\frac{\partial}{\partial r}(ru) + \frac{\partial w}{\partial z} = 0 \qquad (1.4)$$

$$\frac{\partial b}{\partial t} = \beta^2 w \qquad (1.5)$$

Where the symmetry of hurricanes with the azimuthal angle has been used. (u,v,w) are the components of wind field in the radial, tangential and vertical directions, respectively. f is Coriolis parameter, r is radius. b here is the new quantity representing the buoyancy (b=(T-$T_e$)/$T_e$ where $T_e$ is the temperature of environment.) and $\beta^2$ = k/$C_p$-$N^2$ ($N^2$ as usual is the buoyancy frequency and the relation J/$C_p$ = kw has been used in equation (1.5), see Part I for detailed)

Combine (1.3) and (1.4), the solution for w is then given by:

$$w = W_1(r,z)e^{\beta t} + W_2(r,z) \qquad (1.6)$$

As in Part I, it is reasonable to consider the function $W_1(r,z)$ as a separable function of radius and height $W_1(r,z)=H(z)R(r)$. The reason for this separable function is that vertical motion is always equal to zero at both the lower surface and the tropopause. Therefore, the dependence of vertical motion with height can be approximated to be the same everywhere. In Part I, the function $H(z)$ is chosen as: $H(z)=W_0\sin(\lambda z)$, where $\lambda$ is inversely proportional to the standard scale height $H_0$ ($H_0 \approx 10km$) so that the vertical velocity is equal to zero at $z=0$ and $z=H_0$ (see Part I for the explanations of this choice). Also, as in Part I, $W_2(r,z)$ will be neglected. This neglecting means that all vertical motions within the region of radius **a** is assumed to originate from the positive feedback process. As mentioned above, in order to remove the discontinuity of tangential wind and geopotential with radius, it is necessary to replace the top-hat function of $R(r)$ in Part I ($R(r) = 0$ for r<**a** and zero otherwise) by the Gauss-distribution function: $R(r) = e^{-\frac{r^2}{a^2}}$. Therefore, the final solution for the vertical velocity will become:

$$w = H(z)e^{-\frac{r^2}{a^2}}e^{\beta t} \qquad (1.6')$$

It should be remembered that there now no longer have two regions I and II as in the nonlinear theory in Part I since the top-hat function for $R(r)$ has been replaced by the Gauss function. It is therefore unnecessary to solve the system ((1.1) to (1.5)) separately as in Part I. Plugging (1.6') into (1.4) to obtain an equation for radial wind:

$$\frac{1}{r}\frac{\partial}{\partial r}(ru) = -\frac{\partial w}{\partial z} = -\frac{dH}{dz}e^{-\frac{r^2}{a^2}}e^{\beta t} \qquad (1.7)$$

Integrate both sides of (1.7) with respect to radius, it is immediate to have the expression for radial wind as:

$$u = (\frac{dH}{dz}\frac{a^2}{2}e^{-\frac{r^2}{a^2}}e^{\beta t} + C)\frac{1}{r} \qquad (1.8)$$

At r=0, since radial wind must be zero, the constant C must be: $C = -\dfrac{dH}{dz}\dfrac{a^2}{2}e^{\beta \cdot t}$. The final

solution for the radial wind is:

$$u = \frac{dH}{dz}\frac{a^2}{2}e^{\beta \cdot t}(e^{\frac{-r^2}{a^2}}-1)\frac{1}{r} \qquad (1.9)$$

It is not difficult to check that the radial wind will increase linearly with radius for the small

radius and will decrease as a function of inverse radius for the large radius. This behavior of

the radial wind is consistent with that in the nonlinear theory in Part I. This is not surprising

since Gauss function, in some way, is a smooth function of the top-hat function. Therefore, it

is necessary for the radial wind to have behaviors consistent with those in Part I as the radius

approaches the extreme values. The advantage of this Gauss function over top-hat function

for vertical velocity is now clear: the radial wind is automatically continuous without using

any continuous condition at the radius r = **a** as in the nonlinear theory in Part I. It is expected

that this advantage will be also applied for tangential wind and geopential. Substitute (1.6)

and (1.9) into (1.2) to obtain an equation for tangential wind:

$$\frac{\partial v}{\partial t} + Me^{\beta \cdot t}\frac{(e^{\frac{r^2}{a^2}}-1)}{r}\frac{\partial v}{\partial r} + He^{\beta \cdot t}e^{-\frac{r^2}{a^2}}\frac{\partial v}{\partial z} + Me^{\beta \cdot t}\frac{(e^{\frac{r^2}{a^2}}-1)}{r}\frac{v}{r} = -fMe^{\beta \cdot t}\frac{(e^{\frac{r^2}{a^2}}-1)}{r} \quad (1.10)$$

Where $M = \dfrac{dH}{dz}\dfrac{a^2}{2}$. Equation (1.10) is a very complicated PDE, which is hard to solve for

exactly. From the results in the nonlinear theory in Part I, we know that both radial and

tangential wind have the same dependence on radius. Therefore, it is suggestive to seek a

solution for the tangential wind in the same form as the radial wind solution (1.9) (Because

the Gauss-distribution function is a kind of generalization of the top-hat function). We thus

find the solution of tangential wind in the form:

$$v = g(r,z,t)(e^{\frac{-r^2}{a^2}} - 1)\frac{1}{r} \qquad (1.11)$$

Where the function g(r,z,t) is still a function of r. At the large and small radius, g(r,z,t) is expected to be a function of (z,t) only so that solution (1.11) will have the same asymptotic forms as those in the non-linear theory in Part I. If this is the case, it is easy to verify that the tangential wind will have the correct expressions as in Part I when the radius approaches the very large and small values: tangential wind will increase linearly with radius for the small value of r and decrease as a function of inverse radius for the large value of r. Substitute (1.11) into (1.10), with the attention that:

$$\frac{\partial v}{\partial t} = (e^{-\frac{r^2}{a^2}} - 1)\frac{1}{r}\frac{\partial g(r,z,t)}{\partial t} \qquad (1.12a)$$

$$\frac{\partial v}{\partial r} = \{(-\frac{1}{r^2})e^{-\frac{r^2}{a^2}} + \frac{1}{r}e^{-\frac{r^2}{a^2}}(\frac{-2r}{a^2}) + \frac{1}{r^2}\}g + (e^{-\frac{r^2}{a^2}} - 1)\frac{1}{r}\frac{\partial g}{\partial r} \qquad (1.12b)$$

$$\frac{\partial v}{\partial z} = (e^{-\frac{r^2}{a^2}} - 1)\frac{1}{r}\frac{\partial g(r,z,t)}{\partial z} \qquad (1.12c)$$

we will obtain:

$$\frac{(e^{-\frac{r^2}{a^2}} - 1)}{r}\frac{\partial g}{\partial t} + Me^{\beta,t}\frac{(e^{-\frac{r^2}{a^2}} - 1)}{r}\{(-\frac{(e^{-\frac{r^2}{a^2}} - 1)}{r^2} - \frac{2e^{-\frac{r^2}{a^2}}}{a^2})g + \frac{(e^{-\frac{r^2}{a^2}} - 1)}{r}\frac{\partial g}{\partial r}\}$$
$$+ He^{\beta,t}e^{-\frac{r^2}{a^2}}\frac{(e^{-\frac{r^2}{a^2}} - 1)}{r}\frac{\partial g}{\partial z} + Me^{\beta,t}\frac{(e^{-\frac{r^2}{a^2}} - 1)}{r}\frac{(e^{-\frac{r^2}{a^2}} - 1)}{r}\frac{g}{r} = -fMe^{\beta,t}\frac{(e^{-\frac{r^2}{a^2}} - 1)}{r} \qquad (1.13)$$

Where M=$\frac{dH}{dz}\frac{a^2}{2}$. After some simplifications, (1.13) becomes:

$$\frac{\partial g}{\partial t} + Me^{\beta,t}(-\frac{2e^{-\frac{r^2}{a^2}}}{a^2}g + \frac{(e^{-\frac{r^2}{a^2}} - 1)}{r}\frac{\partial g}{\partial r}) + He^{\beta,t}e^{-\frac{r^2}{a^2}}\frac{\partial g}{\partial z} = -fMe^{\beta,t} \qquad (1.14)$$

Consider now two extreme limitations of equation (1.14) at the large and small radiuses.

*At the large radius*

Since exp[-(r/a)²] ≈ 0, (1.14) now becomes:

$$\frac{\partial g}{\partial t} = -fMe^{\beta t} \tag{1.15}$$

The exact solution of (1.15) clearly is:

$$g = -Me^{\beta t}\frac{f}{\beta} + C \tag{1.16}$$

From (1.11), the solution of (1.14) at the large radius is:

$$v_{r,l\arg e} = \{-Me^{\beta t}\frac{f}{\beta} + C\}(e^{-\frac{r^2}{a^2}} - 1)\frac{1}{r} \approx (-Me^{\beta t}\frac{f}{\beta} + C)\frac{1}{r} \tag{1.17}$$

This solution should be compared with the solution (2.33) in Part I. Apparently, both solutions at the large radius are completely the same, which assures that all of the results in the nonlinear theory in Part I can be applied equally well at the large radius. We expect this consistency because the solutions obtained in the nonlinear theory in Part I have been verified with observations. Therefore, any modification in this section, at first, must preserve all the good points that the nonlinear theory in Part I has obtained and the solution (1.17) is perfectly what we are looking for at the large radius.

*At the small radius*

Using approximation: exp(-(r/a)²) ≈ 1-(r/a)², equation (1.14) becomes:

$$\frac{\partial g}{\partial t} + Me^{\beta t}(-\frac{2(a^2 - r^2)}{a^4}g - a^2r\frac{\partial g}{\partial r}) + He^{\beta t}(1 - \frac{r^2}{a^2})\frac{\partial g}{\partial z} = -fMe^{\beta t} \tag{1.18}$$

Keep only terms in (1.18) up to the first order in radius, (1.18) becomes:

$$\frac{\partial g}{\partial t} + Me^{\beta t}(-\frac{2}{a^2}g - a^2r\frac{\partial g}{\partial r}) + He^{\beta t}\frac{\partial g}{\partial z} = -fMe^{\beta t} \tag{1.19}$$

This equation should be compared with equation (2.12) in Part I. Instead of solving this equation directly, it is sufficient to verify that the solution (2.23) in Part I:

$$v_{r.small} = \{G_0 \frac{\sin(\lambda z)}{\tan(\frac{\lambda z}{2})^{\frac{1}{w_0\beta\lambda}}} epx(e^{\beta \cdot t}) + D\sin(\lambda z) - \frac{f}{2}\}r \qquad (1.20)$$

is indeed a particular solution of (1.19). This once again tells us that, at the small radius, equation (1.14) admits a solution which agrees with that in the nonlinear theory in Part I. In order to complete the treatment of the problem, it would be great if equation (1.14) can be solved for all intermediate values of radius, admitting the solutions (1.17) and (1.20) as boundary conditions. However, it is very hard and, fortunately, not very necessary to do that task. Recall that our purpose here is to find a modification such that all conclusions obtained in the nonlinear theory in Part I can be still valid while the discontinuity of tangential wind and geopotential with radius at the later time is removed. Obviously, this purpose is done. Although it is not possible to obtain any explicit solution at all values of radius for tangential wind (and consequently, geopotential), basically we have shown a way to get rid of the discontinuity. This is very important because, from now on, it is sufficient to use the top-hat function for vertical velocity to examine the properties of analytical solutions without any worry about the possible disturbing discontinuities. The reason why the top-hat function is preferred is because exact solutions can be found the most easily in this case. Whenever it is required to remove discontinuities, all we need to do is to replace the top-hat function by the Gauss-distribution function and all conclusions are still preserved.  It should be noted that equation (1.14) will contain many other possible solutions. However, by using the solutions (1.17) and (1.20) as boundary limitations, which are physically meaningful at the large and small radiuses, the continuous physical solution of (1.14) will be unique.

## 3. A nonlinear theory with friction included

All the theories for the early stage of hurricanes developed so far in this series of papers have no friction included. Actually, as explained in part I, the role of friction becomes much less significant in all of the above theories due to the assumption of the existence of a positive feedback between latent heat release and vertical motion. Usually, in a real atmosphere, friction is believed to be the main mechanism for the convergence at low level. This convergence is very important since it provides a source of latent heat for hurricanes. The no-friction assumption may lead to some confusing and the role of friction, therefore, is deserved to be discussed in more detail before we delve into this section.

Normally, the discussions of friction and convergence are mentioned together (and blended into one another). The common thinking about the effect of friction in hurricane development is as follows: Friction will slow down tangential wind and annihilate the balanced tangential wind. So, there will appear an inward component of radial wind and, because of the conservation of angular momentum *(this is not really true due to the presence of friction)*, the inward wind will lead to a strong increase of tangential wind. Also, the convergence toward low center will create an upward motion and provide more moisture to the upper levels. These arguments have an origin from the Ekman convergence problem (e.g. Holton 1993). Therefore, an assumption of no friction may lead to a quick conclusion that there is no convergence and hurricanes cannot exist. However, it turns out that the always side-by-side appearance of friction and convergence conceals another kind of convergence. It is convenient to divide the convergence in hurricanes imaginarily into two kinds. The first is the friction convergence mentioned above. The second kind of convergence is related to a positive feedback process between vertical motion and latent heat and has no relation to friction. The meaning of this feedback convergence is as follows: Suppose now that a positive feedback is triggered at the middle level of the atmosphere, then both temperature and vertical velocity will increase at this level. Because of the impenetrable surface boundary and the stable troposphere, the vertical velocity will be forced to be zero at the surface and tropopause. Since vertical velocity is a continuous function of height, there will have a convergence of wind below (vertical velocity increases with height) and divergence above (vertical velocity decreases with height). This strong convergence (divergence) leads to a larger inward (outward) acceleration of radial wind below (above) and so on. An equivalent view is that a warm core in the feedback region will lead to an expanding column and will lower the surface pressure. A low pressure system will develop and more convergence will occur. Friction plays no role in this feedback process.

In all of the theories for the early stage of hurricane development so far, much of the concentration is on the convergence due to the positive feedback process. Friction convergence has been totally neglected. Basically, these theories support the development of hurricanes in which the role of friction is minor. *No friction does not mean that there is no convergence.* If there is no feedback mechanism, friction alone will spin down a vortex instead of strengthening it. When a positive feedback appears, the roles of this feedback in controlling the convergence will dominate those of friction. Hurricanes, as a whole, are governed by the convergence from the feedback process rather than by the friction convergence. To investigate the role of friction, the nonlinear theory in Part I will be expanded in this section in which friction will be included. It should be careful that all of the variables in the system of the primitive equations are not averaged in any way and, therefore, there is no room for the concept of eddy flux terms in any discussion here (see Part I for the detailed explanations of the role of eddy terms in the primitive equations).

Because it is now known from section 2 that the assumption of the top-hat function for the vertical velocity is not a serious shortcoming (though this function will lead to a discontinuity of tangential wind and geopotential), we will use again this top-hat function for vertical motion so that the exact solutions of the system of the primitive equations can be found most easily. Any discontinuity will be eliminated simply by replacing the top-hat function by the Gauss-distribution function and every conclusions obtained in this section are still valid. All of the starting assumptions in this section will be the same as those in the nonlinear theory in Part I. However, the no-friction assumption, which has been used in Part I, will now be removed. Rewrite the starting system of the primitive equations in the cylindrical coordinate:

$$\frac{\partial u}{\partial t} + u\frac{\partial u}{\partial r} + w\frac{\partial u}{\partial z} - \frac{v^2}{r} = -\frac{\partial \phi}{\partial r} + fv + \kappa\frac{\partial^2 u}{\partial z^2} \tag{2.1}$$

$$\frac{\partial v}{\partial t} + u\frac{\partial v}{\partial r} + w\frac{\partial v}{\partial z} + \frac{uv}{r} = -fu + \kappa\frac{\partial^2 v}{\partial z^2} \tag{2.2}$$

$$\frac{\partial w}{\partial t} = b \tag{2.3}$$

$$\frac{1}{r}\frac{\partial}{\partial r}(ru) + \frac{\partial w}{\partial z} = 0 \tag{2.4}$$

$$\frac{\partial b}{\partial t} = J \tag{2.5}$$

Here we have introduced terms $\kappa\partial^2 u/\partial z^2$ and $\kappa\partial^2 v/\partial z^2$ to represent frictional effects. Note that these friction terms represent only the molecular viscosity because all field variables (u, v, w, etc.) are not subject to the Reynolds average and, therefore, no eddy concept will appear. Following exactly what have been done in Part I, the solutions for the vertical and radial velocity in regions I and II will be given by:

$$w_1(r,z,t) = H(z)e^{\beta \cdot t} \text{ in region I and } w_2(r,z,t) = 0 \text{ in region II} \qquad (2.6)$$

$$u_1 = -\frac{dH}{dz}e^{\beta \cdot t}\frac{r}{2} \equiv Qe^{\beta \cdot t}r \qquad (2.7)$$

$$u_2 = -\frac{1}{2}\frac{dH}{dz}e^{\beta \cdot t}\frac{a^2}{r} \equiv e^{\beta \cdot t}\frac{Q'}{r} \qquad (2.8)$$

Consider first region I where the positive feedback mechanism is effective.

*Region I*

Substitute solutions (2.6) and (2.7) for the vertical and radial wind in region I into (2.2) to obtain an equation for the tangential wind:

$$\frac{\partial v_1}{\partial t} + Qe^{\beta \cdot t}r\frac{\partial v_1}{\partial r} + He^{\beta \cdot t}\frac{\partial v_1}{\partial z} + Qe^{\beta \cdot t}r\frac{v_1}{r} = -Qrfe^{\beta \cdot t} + \kappa\frac{\partial^2 v_1}{\partial z^2} \qquad (2.9)$$

As explained in Part I, it is reasonable to seek the solution $v_1$ in this region in the form: $v_1 = F(z,t)r$. Substitute this solution into (2.9):

$$\frac{\partial F}{\partial t} + He^{\beta \cdot t}\frac{\partial F}{\partial z} + 2Qe^{\beta \cdot t}F = -Qfe^{\beta \cdot t} + \kappa\frac{\partial^2 F}{\partial z^2} \qquad (2.10)$$

or

$$\frac{\partial F}{\partial t} - \kappa\frac{\partial^2 F}{\partial z^2} = -e^{\beta \cdot t}(H\frac{\partial F}{\partial z} + 2QF + Qf) \qquad (2.11)$$

It is not difficult to find a particular solution of (2.11): $F_p = -f/2$, and our task now is to solve the homogeneous solution $F_h$ of (2.11):

$$\frac{\partial F_h}{\partial t} - \kappa\frac{\partial^2 F_h}{\partial z^2} = -e^{\beta \cdot t}(H\frac{\partial F_h}{\partial z} + 2QF_h) \qquad (2.12)$$

As in Part I, the vertical profile of vertical wind is chosen as: H=$W_0$sin($\lambda z$). So, from definition of Q, we have: Q = $\lambda W_0$cos($\lambda z$). Substitute these expressions for H and Q into (2.12):

$$\frac{\partial F_h}{\partial t} - \kappa \frac{\partial^2 F_h}{\partial z^2} = -e^{\beta \cdot t}(W_0 \sin(\lambda z)\frac{\partial F_h}{\partial z} - W_0 \lambda \cos(\lambda z)F_h) \qquad (2.13)$$

The problem now becomes much more complicated because of the appearance of the friction terms, although this term is very small (molecular viscosity coefficient $\kappa$ is usually very small). There no longer exists solutions of the form $F_h$=Cexp($e^{\beta t}$) as in the no-friction theory in Part I. Indeed, it is not hard to prove that, if $\kappa$ is a constant, the only separable solution that we can get from (2.13) will have a form: $F_h = D\sin(\lambda z)e^{-\kappa \lambda^2 t}$, which will decrease with time and it is not the solution that we are expecting for a developing system. The expected solution of (2.13) must be not only a growing solution abut also a continuous function of $\kappa$ such that, when the viscosity coefficient is equal to zero, this solution needs to approach the solution obtained in the no-friction nonlinear theory in Part I. Equation (2.13) illustrates very clearly how much a small quantity when added can effect to the final exact solution. Normally, the scaling evaluation is used and one simply ignores a small molecular frictional term. This small term may have a small effect to the numerical calculation but it can change substantially the functional form of the exact solution. Sometimes, this small adding term will create a bifurcation point and exact solutions may behave totally different. Consequently, the values obtained from numerical calculation needs to have a very special attention.

However, if all field variables are not Reynolds averaged as stated above, this viscosity coefficient $\kappa$ will be very small and it represents just a molecular friction. In order to have a more meaningful effect of friction, it is necessary to apply the Reynolds-averaging operator to transform the true variables to Reynolds-averaged variables so that friction will represent eddy processes. The system of the primitive equations does not change under this averaging process except there appear some eddy terms (see e.g. Holton 1993, chapter 5). We will always understand that if friction terms represent eddy processes, all variables are Reynolds averaged. In this case, the appearance of eddy processes will make problem become extremely complicated since it is impossible to know a true parameterization for the small-scale processes. For the purpose of theoretically investigating, it is desired to use some tentative functional forms for the eddy terms to study the characteristics of the exact solutions. Consider now the case in which the eddy coefficient $\kappa$ is a function of time as follows: $\kappa = \kappa_0 \exp(\beta t)$. The motivation for this choice comes from the observations and previous works that the eddy coefficient is a function of wind speed. If a hurricane is growing, wind speed will increase with time and, therefore, this eddy coefficient $\kappa$ will depend on time. This time dependence may be not the actual case but, at least, it provides us some insights into the effect of friction and makes problem solvable. With this assumption, (2.13) now becomes:

$$\frac{\partial F_m}{\partial t} - \kappa_0 e^{\beta t} \frac{\partial^2 F_m}{\partial z^2} = -e^{\beta t} (W_0 \sin(\lambda z) \frac{\partial F_m}{\partial z} - W_0 \lambda \cos(\lambda z) F_m) \qquad (2.14)$$

Because the RHS of (2.14) is a function of the form: $(H \frac{\partial F_m}{\partial z} - \frac{dH}{dz} F_m)$, it may be tempted to find a solution of (2.14) in the form: $F_m = HG(t) = \sin(\lambda z)G(t)$. However, after plugging this solution into (2.14) and doing some simplifications, it will be easy to find that the final solution will decrease with time and it is not what we expect for hurricane development. A quick inspection will tell us that a growing solution of (2.14) can be found in the form: $F_m = \exp(e^{\beta \cdot t})G(z)$. Plug this solution into (2.14):

$$\frac{G}{\beta} - \kappa_0 \frac{d^2 G}{dz^2} = -(W_0 \sin(\lambda z)\frac{dG}{dz} - W_0 \lambda \cos(\lambda z)G) \qquad (2.15)$$

or

$$\kappa_0 \frac{d^2 G}{dz^2} - W_0 \sin(\lambda z)\frac{dG}{dz} + (W_0 \lambda \cos(\lambda z) - \frac{1}{\beta})G = 0 \qquad (2.16)$$

Once again, (2.16) is not easy to solve but the important point is that we now have G(z) as a function of z only and totally, the explicit solution will depend on time as an exponential function of exponential of time. The only change compared with the case of no friction in Part I is that the vertical profile of wind will be modified. The role of friction in this case is to re-arrange the dependence of tangential wind with height. The final solution for tangential wind in region I can be written symbolically as

$$v_1 = C\{\exp(e^{\beta \cdot t}) - \frac{f}{2}\}G(z)r \qquad (2.17)$$

where the function G(z) is still unknown.

_Region II_

Substitute (2.6) and (2.8) into equation (2.2) to get an equation for the tangential wind in region II:

$$\frac{\partial v_2}{\partial t} + \frac{Q'}{r}e^{\beta \cdot t}\frac{\partial v_2}{\partial r} + \frac{Q'}{r}e^{\beta \cdot t}\frac{v_2}{r} = -f\frac{Q'}{r}e^{\beta \cdot t} + \kappa\frac{\partial^2 v_2}{\partial z^2} \qquad (2.18)$$

Assuming the form of $v_2$ is: $v_2 = G(z,t)/r$, substitute this solution into (2.18):

$$\frac{1}{r}\frac{\partial G}{\partial t} - \frac{Q'}{r}e^{\beta \cdot t}\frac{G}{r^2} + \frac{Q'}{r}e^{\beta \cdot t}\frac{G}{r^2} = -f\frac{Q'}{r}e^{\beta \cdot t} + \kappa\frac{1}{r}\frac{\partial^2 G}{\partial z^2} \qquad (2.19)$$

Since (2.19) is applied for the whole region II, where radius is greater than zero, it follows that:

$$\frac{\partial G}{\partial t} = -fQ'e^{\beta_{J}t} + \kappa \frac{\partial^2 G}{\partial z^2} \tag{2.20}$$

With Q' defined from (2.8) and H=$W_0$sin($\lambda z$), we have: Q'=$-\lambda W_0$cos($\lambda z$)$a^2$/2. Substitute this expression for Q' into (2.20):

$$\frac{\partial G}{\partial t} = \frac{1}{2}\lambda W_0 f \cos(\lambda z)a^2 e^{\beta_{J}t} + \kappa \frac{\partial^2 G}{\partial z^2} \tag{2.21}$$

This is a familiar diffusion equation. Unlike in region I, it is now possible to find the analytical solution of (2.21) even in the case the viscosity coefficient $\kappa$ is a constant. Consider a particular solution of (2.21) in the form: G = $e^{\beta t}$Z(z) and plug into (2.22):

$$\beta e^{\beta_{J}t}Z = -fQ'e^{\beta_{J}t} + \kappa e^{\beta_{J}t}\frac{d^2 Z}{dz^2} \tag{2.22}$$

or

$$-\kappa \frac{d^2 Z}{dz^2} + \beta Z = \frac{1}{2}fW_0 a^2 \lambda \cos(\lambda z) \tag{2.23}$$

From (2.23), it is simple to get the solution for Z as:

$$Z = \frac{fW_0 a^2 \lambda}{2(\kappa \lambda^2 + \beta)}\cos(\lambda z) \tag{2.24}$$

This is not the final solution for (2.21) yet because we have to consider one more thing: the homogeneous solutions of equation (2.21). It is easy now to find that there are an infinitive number of homogeneous solutions. These homogeneous solutions can be divided into three families, based on their time dependence: a family of time-decreasing solutions, a constant value and a family of time-increasing solutions. The final homogeneous solution requires a superposition of all of these families. This superposition leads to an infinitive solution, which is not acceptable for realistic cases. To get rid of this difficulty, we need to use a condition which has not been used before: *a continuous dependence of the solution of equation (2.21) on κ*. It means that when κ approaches zero, the solution of (2.21) must be identical with that obtained in the no-friction nonlinear theory in Part I (solution with κ=0). Using this condition, the only acceptable homogeneous solution of (2.21) is a constant value. Therefore, the final solution for the tangential wind in the region II is:

$$v_2 = (Ze^{\beta t} + Z_h)\frac{1}{r} = (\frac{fW_0 a^2 \lambda}{2(\kappa \lambda^2 + \beta)}\cos(\lambda z)e^{\beta t} + C)\frac{1}{r} \qquad (2.25)$$

Compare this solution with (2.33) in the nonlinear theory in Part I, it is immediate to realize that if κ is equal to zero, the solution for the tangential wind in region II given by (2.25) is consistent with that obtained from the no-friction nonlinear theory in Part I. For the case in which κ is a constant, friction will represent the molecular viscosity only and κ is very small . Therefore, the difference between solution (2.25) and the solution obtained in Part I will be insignificant.

The effect of friction is the most effective only if the Reynolds-averaged variables are considered as mentioned in few paragraphs above. Consider now a situation in which the eddy coefficient κ will change with time as: κ=κ₀exp(βt) as investigated in region I. In this case, it is not difficult to realize that the solution of (2.21) will have a form: G=T(t)cos(λz). Substitute this solution into (2.21) we have:

$$\cos(\lambda z)\frac{dT}{dt} = \frac{1}{2}\lambda W_0 f \cos(\lambda z)a^2 e^{\beta t} - \kappa_0 e^{\beta t}T\lambda^2 \cos(\lambda z) \qquad (2.26)$$

Because (2.21) has to be valid at all time and level, it must follow that:

$$\frac{dT}{dt} = \frac{1}{2}\lambda W_0 f a^2 e^{\beta t} - \kappa_0 T\lambda^2 e^{\beta t} \qquad (2.27)$$

Since the homogeneous solution of (2.27) is: $T_h = epx(-\frac{\kappa_0 \lambda^2}{\beta} e^{\beta \cdot t})$, the solution of (2.27)

will have the form $T = X(t) epx(-\frac{\kappa_0 \lambda^2}{\beta} e^{\beta \cdot t})$. A simple substitution and integration will give

us the functional form of X(t):

$$X(t) = \frac{\lambda W_o f a^2}{2} \int_a^t epx(\beta y + \frac{\kappa_0 \lambda^2}{\beta} e^{\beta \cdot y}) dy \qquad (2.28)$$

Using the continuous dependence of solution of (2.21) on κ as explained above, the only

acceptable homogeneous solution of (2.21) is a constant. Finally, the complete solution for

tangential wind in region II will be given by:

$$v_2 = \{\frac{\lambda W_o f a^2 \cos(\lambda z) epx(-\frac{\kappa_0 \lambda^2}{\beta} e^{\beta \cdot t})}{2} \int_\zeta^t epx(\beta y + \frac{\kappa_0 \lambda^2}{\beta} e^{\beta \cdot y}) dy + C\} \frac{1}{r} \qquad (2.29)$$

where ζ is a constant of integration. A quick inspection will show that, if $\kappa_0$ is equal to zero,

(2.29) is indeed consistent with solution (2.33) in the no-friction nonlinear theory in Part I.

Once again, we found that the tangential wind in region II increases approximately as an exponential function of time, which is much slower than tangential wind in region I (Note that there is a substantial cancellation of time-dependent terms in (2.29)). This result is also obtained in the no-friction nonlinear theory in Part I.

The final task of finding geopotential for each region is nothing special and will not

be mentioned here. The appearance of the friction terms does not change much the

conclusions about geopotential obtained in Part I except some minor modifications, e.g. the

solution for geopotential in region I now becomes:

$$\phi_1 = \Phi_o - (\beta e^{\beta \cdot t} Q + e^{2\beta \cdot t} Q^2 + H e^{2\beta \cdot t} \frac{dQ}{dz} - K^2 - fKr - \kappa \frac{d^2 Q}{dz^2} e^{\beta \cdot t}) \frac{r^2}{2} \qquad (2.30)$$

Where K= $\{C\exp(e^{\beta t}) - \dfrac{f}{2}\}G(z)$ (This expression for K is applied only to the case in which all variables are Reynolds-averaged and eddy coefficient κ is given by κ=κ₀exp(βt)). It is not possible to have an exact solution the analytical solution of tangential wind in region I because function G(z) is unknown. However, it is reasonable to have all conclusions about the behaviors of geopotential in region I as obtained so far in Part I. This is because friction contributes only: 1) a term which increases as an exponential function of time (the last term inside the bracket in the RHS of (2.30)), which increase much slower than the K-terms and 2) changing the functional form of height dependence (the unknown function G(z)). Therefore, the K-terms in the RHS of (2.30) are still an exponential function of exponential of time and will be the dominated terms at later time. The conclusions for the geopotential field in region II obtained in the no-friction nonlinear theory in Part I are also valid here.

## 4. Unresolved problems

At this point, it can be easily seen that all the theories developed so far in Part I and II still have a very disturbing result: a nonstop growing of all variables. This is because we have used a positive feedback which is completely independent on any internal mechanism. A question is "What prevent a hurricane from growing endlessly while it has an unlimited source of moisture at the ocean surface?" Several reasons can be listed, such as the latent heat release will raise the temperature of environment in (the)region I so that the buoyancy forcing will reach an asymptotic value instead of increasing exponentially with time ($\beta^2$ is a function of time, not a constant as assumed so far), the nonlinear dependence of friction on wind speed (which makes friction become important at later time and the energy lost due to friction can not be simply neglected). Until now, none of these damping effects was considered. The results in this work, as mentioned in Part I, should be confined to the first 6-12 hours of hurricane lifetime. Beyond this period, all solutions will be inaccurate.

In section 3, there are some investigations into the role of friction at the early stage of hurricane development, which was ignored in Part I. It was commented that there are basically two kinds of convergence: one is due to friction and the other is due to a feedback process between latent heat release and vertical motion. According to the theories developed so far in Part I and II, the role of the feedback process in controlling the convergence is so influent that friction, when included, does not change significantly all the conclusions obtained in the no-friction theory in Part I as seen in section 3. It is worthy now to have some more explanations about this separation between two kinds of convergence. In an axis-symmetric hurricane, the variation of tangential wind with the azimuthal angle, $\partial v / \partial \lambda$, is always assured to be zero and all of the convergence/divergence will be decided only by the vertical and radial motions. In the presence of friction, tangential wind will be slowed down (but the axis-symmetry still guarantees that $\partial v/\partial \lambda = 0$) and there will appear an inward acceleration of radial wind. This acceleration creates a convergence at low level and, consequently, vertical motion will follow. In a typical atmosphere, such friction-induced secondary circulation will spin down a vortex (Holton, 1993). However, in hurricane conditions, such a circulation may create an unknown feedback and, so far, this friction feedback has not been included. Note that this friction feedback is different from the feedback employed in this series of papers, which is a relationship between vertical motion and latent heat. There appears now a difficult question: "how can we distinguish between these feedbacks clearly". Actually, this work has answered partly this question by pointed out that, under only the assumption of the feedback between latent heat and vertical motion, the dynamical properties of hurricanes have been captured well. This fact shows implicitly that the feedback between vertical motion and latent heat is dominant. It should be mentioned that, though friction was included in section 3, the feedback is still between latent heat and vertical motion only and the role of friction feedback is still open.

As in Part I, nothing is mentioned about the eye of hurricanes because all of our concentration is just on the early stage of hurricane development. However, with the explicit solutions obtained in the nonlinear theory in Part I and this paper, we can say something about the appearance of an eye of a hurricane. It is possible to imagine this eye-forming process of a hurricane as follows: When a positive feedback is effective within some region, vertical wind will increase as an exponential function of time in this region. This strong vertical motion, or equivalently the strong vertical advection, will generate a large source term in the tangential wind equation. As a consequent, the tangential wind in this region will increase very fast with time and create a parabolic shape of geopotential (in the same way as we stir robustly a water surface in a glass). Here, we have a situation: wind field decides the shape of geopotential rather than inversely. (This is reasonable since, at small scale, the atmosphere tends to have mass field adjusted to wind field rather than inversely). As time passes, this geopotential surface is lower and lower and an eye tends to form at later time. However, this picture is still far from observations since vertical velocity, according to the theories in Part I and II, increases exponentially everywhere inside region I, which is not realistic as observed. However, at least, the theories developed in Part I and this paper provide a good picture of hurricane dynamics at the early stage of hurricane lifetime.

## 5. Discussion and conclusion

In this second part, my first purpose is to remove the discontinuity of the tangential wind and geopotential with radius, which appears in the analytical solutions for the early developing stage of a hurricane obtained in the non-linear theory in Part I. By replacing the top-hat function by the Gauss-distribution function for vertical velocity, it is then found that the discontinuity has been eliminated satisfactorily without changing all conclusions obtained in the nonlinear theory in Part I. Although it has not been possible to find all solutions completely explicitly in the case of using the Gauss-distribution function, this improvement is helpful because it allows us to continue using the top-hat function for vertical velocity without any worry about the appearance of discontinuity. The reason we need to use the top-hat function for investigation is because this is the case in which the analytical solutions of the primitive equations can be found the most easily.

In section 3, the influences of friction were taken into account, which has been neglected in the nonlinear theory in Part I. If the primitive equations are not subject to Reynolds-averaging process, then these frictional effects represent just the molecular viscosity and, therefore, they are of no interest to us. In order to have a more significant meaning of the influence of friction, all variables are then Reynolds averaged so that the frictional terms will contain eddy processes. In this case, it was found that all conclusions obtained in the no-friction theory in Part I are still valid. The influences of friction are thus minor.

It is perhaps surprising that, by assuming just a positive feedback between vertical motion and latent heat, it is then possible to find all solutions of the system of primitive equations explicitly as seen in Part I and II. As pointed out in Part I, this is because the system of the primitive equations has been divided into two closed subsystems: one for radial and vertical velocities and the other for tangential wind and geopotential. Under the assumption of feedback process, it is then easy to obtain the explicit expression for vertical motion with time. Using the continuity equation, we then found radial wind and then tangential wind and finally geopotential. A little scrutiny will show that, in this process of finding solutions, the role of the feedback process is only apparent in determining the time dependence of vertical motion. If suppose now that, by some way, we have an explicit expression for time dependence of the vertical motion, then it is completely unnecessary to employ the feedback assumption and we are still able to have all solutions obtained so far. There are two properties that make the primitive system becomes particularly solvable for hurricane phenomenon. The first property is the existence of the impenetrable surface and the tropopause. These boundaries specify an initial prescribed functional form for vertical motion such that the vertical velocity must be zero at these boundaries, regardless of type of motion. Several functional forms for the dependence of the vertical velocity on height can be used but the final results would not be changed much because the primitive equations are the first-order partial differential equations. These boundaries will give us an explicit function for the height-dependent part of the vertical velocity in any situation. The second property is the axis-symmetry of hurricanes. *These two properties make the system of the primitive equations become immediately analytically solvable when vertical velocity (or radial wind) is known a priori*. It is possible to obtain all analytical solutions of the system of the primitive equations by applying only the two above properties without any aiding from the feedback assumption. To see this point clearly, let us consider the following arguments, which are based entirely on these two above properties and also give us all solutions as in Part I and II:

- For the problem of investigating an initial stage of hurricane development, vertical velocity is expected to be an increasing function of time. So, we can write w=T(t)Z(z)R(r), where T(t) is an increasing function of time, assumed to be T(t)= $e^{\beta t}$

- Because of the impenetrable boundaries, the functional form of Z can be chosen to be Z=sin($\lambda$z)

- Assume the vertical motion is large within a region of radius r=**a** and small outside, then we can choose: R(r)= $e^{\frac{-r^2}{a^2}}$ (or top-hat function)

The reason why a separable function for the vertical motion is chosen is because the height-dependence of vertical velocity can be regarded as the same everywhere (due to the first-order derivative of the system of the primitive equations). With these three simple arguments, we have showed that, basically, it is possible to have an initialization for the development of a vortex without any aid from the feedback assumption. Follow completely the same as in Part I, all the dynamical characteristics of hurricane development can be recovered. This is a remarkably simple route to find exact solutions for the primitive equations. This route is of course arbitrary but its simplicity is very amazing. If one attempts to find first tangential wind or geopotential and next vertical and radial wind, it will become a nearly impossible task.

There now appear questions, such as: What is the meaning of the order of solving a system of equations? Can this order represent the true evolution of a hurricane or it just a mathematical trick? This question is actually not trivial. In the experiment with a water surface, it is seen clearly that the pattern of the water surface is decided by the velocity field, (which is given by an equation: $\partial p/\partial r = v^2/r$). But in a synoptic motion, this order is inverted as $v^2/r = \partial p/\partial r$ (note the order of writing this equation) and the meaning, obviously, is also different: the velocity field will be decided by the pressure field. The different way to solve a system of equations tells us a different process in nature.

# Reference

Challa, M., R. L. Pfeffer,Q. Zhao, S. W. Chang, 1998: Can Eddy Fluxes Serve as a Catalyst
for Hurricane and Typhoon Formation?. *J. Atmos. Sci.* **55**, 2201-2219

Craig, G. C. and S. L. Gray, 1996; CISK or WISHES as the Mechanism for tropical Cyclone
Intensification, *J. Atmos. Sci*., **53**, 3528-3540

Emanuel, K. A., 1986: An air-sea Interaction Theory for Tropical Cyclone. Part I: Steady
state maintenance, *J. Atmos. Sci*., **43**, 585-604

________, 1988: The Maximum Intensity of Hurricanes, *J. Atmos. Sci*., **45**,
1143-1155

________, 1999: Thermodynamic control of hurricane intensity, *Nature*, 401, 665-669

Holland, G .J.: mature structure and Structure changes, Chapter 2, (unknown source)

________, 1980: An Analytical Model of the Wind and Pressure Profiles in Hurricanes,
*Mon. Wea. Rev.,* **108**, 1212-1218

________, 1997: The maximum Potential Intensity of Tropical Cyclones, *J. Atmos. Sci*.
**54**, 2519-2541

Kieu, C. Q., 2004: An Analytical Theory for the Early Stage of the Development of
Hurricanes: Part I, available in website: www.arxiv.org

Kuo, H. C., R. T. Williams and J-H Chen, 1999: A Possible Mechanism for the Eye Rotation
of Typhoon Herb, *J. Atmos. Sci*. **56**, 1659-1673

Kwon, H. J., S-H Won, M-H Ahn, A-S Suh and H-S Chung, 2002: GFDL-Type Typhoon
Initialization in MM5, *Mon. Wea. Rev.,* **130**, 2966-2974

Liu, Y., Dalin Zhang and M. K. Yau, 1999: A multiscale Numerical Study of Hurricane
Andrew. Part II: Kinematics and Inner-core Structures, *Mon. Wea. Rev,* **127**,

2597-2616

McBride, J. L. 1981: Observational Analysis of Tropical Cyclone Formation. Part I: Basic

Description of Data Sets, *J. Atmos. Sci.*, **38**, 1117-1131

_______and R. Zehr, 1981: Observational Analysis of Tropical Cyclone Formation.

Part II: Non-Developing versus Developing Systems, *J. Atmos. Sci.*, **38**,

1132-1151

Moller, J. D. and L. J. Shapiro, 2002: Balanced Contributions to the Intensification of

Hurricane Opal as Diagnosed from a GFDL Model forecast. *Mon. Wea. Rev.*, **130**,

1866-1881

Montgomery, M. T. and B. F. Farrell, 1993: Tropical Cyclone Formation, *J. Atmos. Sci.*, **50**,

285-310

Nolan, D. S. and M. T. Montogomery, 2002: Nonhydrostatic, Three_dimensional

Perturbations to balanced, Hurricane-like Vorties, Part I: Linearized Formation,

Stability, and Evolution, J. Atmos. Sci., **59**, 2989-3020

Nguyen, M. C., Hongyan Zhu, R. K. Smith, W. Ulrich, 2001: A Minimal axi-symmetric

Tropical Cyclone Model**,** *Q. J. R. Meteo. Soc.*, **128**, 1-20

Palmen E., and C. W. Newton, 1969: Atmospheric Circulation Systems, *Academic*

*Publisher*,   603p

Pu, Z-X. and S. A.  Braun, 2001 : Evaluation of bogus Vortex Techniques with Four-

Dimensional Variational Data Assimilation, *Mon. Wea. Rev.*, **129**, 2023-2039

Rodgers, E.B., W. S.Olson, V. M. Karyampudi and H. F.  Pierce, 1998: Satellite-Derived

Latent Heating Distribution and Environmental Influences in Hurricane Opal (1995).

*Mon. Wea. Rev.,* **126**, 1229-1247

Shapiro, L. J. and M. T. Montgomery, 1993: A three-dimensional balance Theory for

Rapidly Rotating Vorties., *J. Atmos. Sci*., **50**, 3322-3335

Smith, R. K., 1980: Tropical Cyclone Eye Dynamics. *J. Atmos. Sci*., **37**, 1227-1232

Wang, Xingbao and Dalin Zhang, 2003: Potential Vorticity Diagnosis of a Simulated
Hurricane. Part I: Formulation and Quasi-Balanced Flow, *J. Atmos. Sci*., **60**,
1593-1607

Willoughby, H. E., 1979: Forced Secondary Circulations in Hurricanes, *J. Geophys. Res*.,
**84**, 3172-3183

_______ 1990: Temporal Changes of the Primary Circulation in tropical Cyclone,
*J. Atmos. Sci*., **47**, 242-264

_______, J. A. Clos and M. G. Shoreibah, 1982: Concentric Eye Walls, Secondary
Wind Maxima, and The Evolution of the Hurricane Vortex, *J. Atmos. Sci*., **39**,
395-411

Zhang, D-L, Y. Liu and M. K. Yau, 2000: A multiscale Numerical Study of Hurricane
Andrew. Part III: Dynamically Induced Vertical Motion, *Mon. Wea. Rev*, **128**,
3772-3788

_______, Y. Liu and M. K. Yau, 2001: A multiscale Numerical Study of Hurricane
Andrew. Part IV: Unbalanced Flows, *Mon. Wea. Rev*, **129**, 92-107

_______, Y. Liu and M. K. Yau, 2002: A multiscale Numerical Study of Hurricane
Andrew. Part V: Inner-core thermodynamics, *Mon. Wea. Rev*, **130**, 2745-2763

Zhu, Hongyan, R. K. Smith, W. Ulrich, 2001: A Minimal Three-Dimensional Tropical
Cyclone Model**,** *J. Atmos. Sci*., **58**, 1924-1944